\documentclass[twocolumn,aps,prl,superscriptaddress,nofootinbib,showpacs,floatfix]{revtex4-1}
\RequirePackage{lineno}
\usepackage{url}
\usepackage{cancel}
\usepackage[colorlinks,linkcolor=blue,citecolor=blue,filecolor=black,urlcolor=blue]{hyperref}
\usepackage[normalem]{ulem} 
\usepackage{epsfig,graphics}
\usepackage{graphicx}
\usepackage{dcolumn}
\usepackage{bm}
\usepackage[usenames]{color}
\usepackage{ulem}
\usepackage{amssymb}
\usepackage{xspace} 
\usepackage{amsmath}
\usepackage{multirow}
\usepackage{float}
\usepackage{harpoon}
\usepackage{MnSymbol}
\usepackage{appendix}
\usepackage{color}


\makeatletter 
\@addtoreset{equation}{section}
\makeatother  

\newcommand{\snn}{\mbox{$\sqrt{s_{\mathrm{NN}}}$}}

\newcommand{\pT} {p_{\mathrm{T}}}

\newcommand{\lr}[1]{\left\langle #1\right\rangle}

\newcommand{\Dphi}{\mbox{$\Delta \phi$}\xspace}
\newcommand{\Deta}{\mbox{$\Delta \eta$}\xspace}
\newcommand{\nch}{N_{\mathrm{ch}}}

\newcommand{\dau}{\mbox{$d$$+$Au}\xspace}
\newcommand{\oo}{\mbox{O+O}\xspace}

\let\oldalign\align
\let\oldendalign\endalign
\renewenvironment{align}{\linenomathNonumbers\oldalign}{\oldendalign\endlinenomath}

\graphicspath{{stardata/}{./}}

\usepackage{CJK}
\hfuzz=\maxdimen
\begin{document}
\title{Engineering the shapes of quark-gluon plasma droplets by comparing anisotropic flow in small symmetric and asymmetric collision systems}
\author{The STAR Collaboration}
\begin{abstract}
The observation of collective flow phenomena in small collision systems challenges our understanding of quark-gluon plasma (QGP) formation and evolution. This complexity lies in the initial geometries, which are influenced by both nucleon configuration and subnucleonic fluctuations, introducing uncertainties in interpreting flow patterns. We disentangle these contributions through comparative measurements of elliptic ($v_2$) and triangular ($v_3$) flow in asymmetric $d$+Au and symmetric $^{16}$O+$^{16}$O collisions at $\sqrt{s_{NN}}=200$ GeV, which produce medium of comparable sizes but with vastly different initial geometries. The larger $v_2$ in $d$+Au reflects its dominant elliptic geometry, while the similar $v_3$ in both systems is better explained by considering subnucleonic fluctuations. These contrasting flow patterns are quantitatively described by a state-of-the-art hydrodynamic model tuned to large-system Au+Au data, indicating efficient transformation of initial geometries to final-state anisotropies. These results provide evidence for droplet formation in small systems with transport properties that are similar to those observed in large collision systems, consistent with QGP-like behavior.
\end{abstract}
\maketitle

{\bf Introduction.} 
High-energy collisions of large nuclei at the relativistic heavy ion collider (RHIC)~\cite{STAR:2005gfr,PHENIX:2004vcz} and the large hadron collider (LHC)~\cite{Heinz:2013th} create a hot, dense state of nuclear matter consisting of quarks and gluons--the quark-gluon plasma (QGP). The QGP undergoes rapid hydrodynamical expansion driven by large pressure gradients, transforming initial spatial anisotropies characterized by eccentricities $\varepsilon_n$ into measurable momentum anisotropies in the azimuthal particle distribution \mbox{$dN/d\phi~\propto1+2\sum v_n\cos(n(\phi-\Psi_n))$}, where $v_n$ and $\Psi_n$ represent the amplitude and orientation of the $n^{\mathrm{th}}$-order anisotropic flow. In the hydrodynamic picture, the dominant flow harmonics, elliptic flow $v_2$ and triangular flow $v_3$ arise from approximately linear responses to the ellipticity $\varepsilon_2$ and triangularity $\varepsilon_3$, respectively~\cite{Niemi:2012aj}. Systematic studies of $v_n$ and their event-by-event probability distributions $p(v_n)$ have established the QGP as a strongly interacting, nearly inviscid fluid~\cite{Busza:2018rrf}.

Understanding the formation criteria of QGP via small collision systems represents a key frontier of research~\cite{Grosse-Oetringhaus:2024bwr,Noronha:2024dtq}. Measurements in $pp$~\cite{CMS:2010ifv,ATLAS:2015hzw,CMS:2016fnw} and small asymmetric collisions such as $p$+Pb~\cite{CMS:2012qk,ALICE:2012eyl,ATLAS:2012cix}, $p$+Au, $d$+Au, $^{3}$He+Au~\cite{PHENIX:2013ktj,PHENIX:2018lia,STAR:2022pfn,STAR:2023wmd} have revealed substantial $v_n$ coefficients that exhibit dependencies on transverse momentum ($\pT$), pseudorapidity ($\eta$), and particle species strikingly similar to those observed in large collision systems like Au+Au and Pb+Pb~\cite{STAR:2004jwm,PHENIX:2006dpn,ATLAS:2012at,ALICE:2016cti,CMS:2013wjq}. However, while $v_n$ in large systems is well-established as a hydrodynamic response to initial geometric anisotropies, its origin in small systems remains debated~\cite{Schenke:2021mxx,Noronha:2024dtq}. 

The central question is whether these small systems can generate a medium with sufficient interactions to form QGP-like matter, or whether alternative mechanisms such as initial-state momentum correlations~\cite{Schenke:2019pmk} can explain the observed collective behavior. Although any reasonable explanation of the $v_n$ across small systems appears to require a hydrodynamical component~\cite{Schenke:2021mxx,Zhao:2022ugy,Wu:2023vqj}, significant uncertainties in the initial geometries of these systems limit our ability to constrain contributions from alternative mechanisms, hampering confirmation of the formation of a QGP-like matter. 

A fundamental challenge in small asymmetric systems stems from the complex interplay between nucleon configuration and subnucleonic fluctuations that govern the initial collision geometry~\cite{Schenke:2021mxx,STAR:2023wmd,Huang:2025cjm}. The subnucleonic degrees of freedom in projectile nuclei ($p$, $d$, $^3$He) play a dominant role in defining the initial geometry, yet modeling these fluctuations carries substantial theoretical uncertainties~\cite{Mantysaari:2016ykx,Loizides:2016djv}. These fluctuations are essential for explaining the large $v_n$ values observed in $pp$ and $p$+Pb collisions at the LHC~\cite{Schenke:2014zha,Moreland:2018gsh} and for interpreting the ordering of the $v_3$ data in $p/d/^{3}$He+Au collisions at RHIC~\cite{PHENIX:2018lia,STAR:2022pfn, STAR:2023wmd}. However, current experimental data lack the control needed to quantify the relative influence of these competing geometrical effects.

We address this challenge by comparing asymmetric systems with small symmetric collisions, specifically $^{16}$O+$^{16}$O. This comparison offers several advantages: 1) Due to oxygen's larger mass number than $p/d/^3$He nuclei, $^{16}$O+$^{16}$O collisions are expected to be less sensitive to subnucleonic fluctuations than $p/d/^3$He+Au collisions~\cite{Huang:2025cjm}. 2) These two system types are expected to create matter with similar final-state collective responses, making $v_n$ ratios exquisitely sensitive to differences in their initial geometries~\cite{Giacalone:2021uhj}. 3) Since nucleon configurations in light nuclei are well-described by {\it ab initio} approaches~\cite{Hergert:2020bxy,Ekstrom:2022yea}, this comparison enables us to vary and disentangle the relative contribution from subnucleonic fluctuations in a theoretically tractable manner.

\begin{figure}
\includegraphics[width=0.8\linewidth]{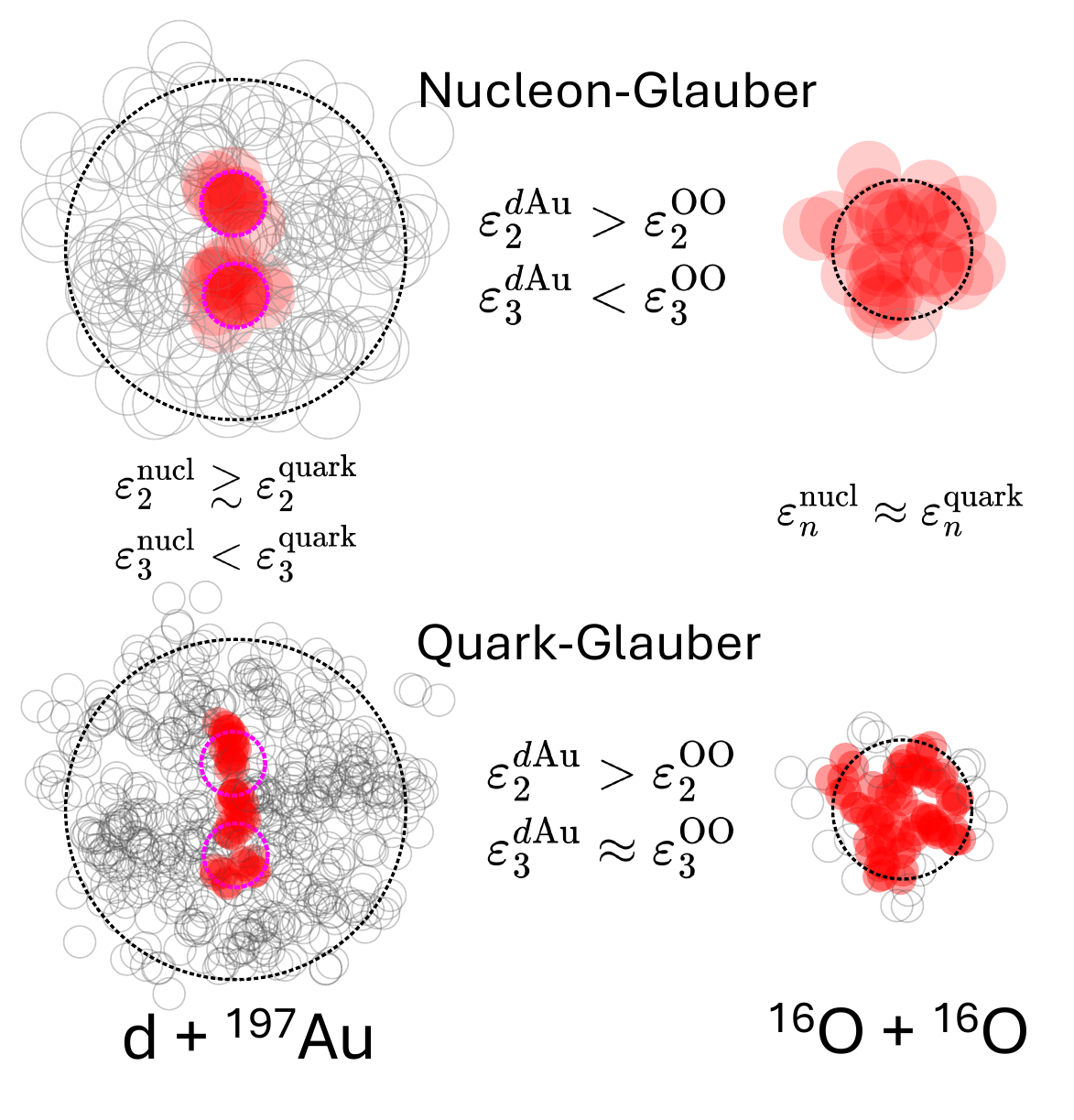}
\caption{\textbf{Contrasting initial geometry in $d$+Au and O+O collisions.} Illustration of one ultra-central $d$+Au (left) or O+O (right) collision for nucleon-Glauber model (top) and quark-Glauber model where each nucleon is replaced with three constituent quarks (bottom). Red blobs indicate participating nucleons (top) or quarks (bottom). The typical ordering of eccentricities $\varepsilon_n$ between the systems and models is shown~\cite{Huang:2025cjm}. Black-dotted circles indicate Wood-Saxon radii for Au (6.38 fm) and O (2.61 fm). Purple-dotted circles show the locations of two nucleons of the deuteron.}
\label{fig:1}
\end{figure}

We implement this comparative approach by leveraging new high-statistics $d$+Au and O+O data from the STAR experiment. As illustrated in Fig.~\ref{fig:1}, the elongated deuteron and near-spherical $^{16}$O nucleus create very different initial geometries that serve as sensitive probes of the produced medium. Glauber model studies~\cite{PHENIX:2018lia,Huang:2025cjm} demonstrate that $\varepsilon_2^{d\mathrm{Au}}$ is dominated by positions of the two nucleons in the deuteron and exhibits weak dependence on subnucleonic fluctuations, while $\varepsilon_3^{d\mathrm{Au}}$ is modestly enhanced by such fluctuations. In contrast, both $\varepsilon_2^{\mathrm{OO}}$ and $\varepsilon_3^{\mathrm{OO}}$ show modest sensitivity to subnucleonic fluctuations, as well as nucleon–nucleon correlations within the oxygen nucleus~\cite{Furutachi:2007vz,Epelbaum:2013paa,Lim:2018huo,Rybczynski:2019adt,Giacalone:2024luz,Zhao:2024feh,Zhang:2024vkh,Loizides:2025ule}.

The contrasting geometric properties between \dau and O+O collisions provide a powerful test on the final-state collective responses. This is accomplished through measurements of $v_2$ and $v_3$ and comparisons to $\varepsilon_2$ and $\varepsilon_3$. The properties of $p(v_n)$ are characterized through its moments $\lr{v_n^2}$ and $\lr{v_n^4}$, which are experimentally accessible via two- and four-particle correlations:
\begin{align}\nonumber
 v_n\{2\} &= \sqrt{\lr{v_n^2}}\;,\\\label{eq:1}
 v_n\{4\} &= \left(2\lr{v_n^2}^2-\lr{v_n^4}\right)^{1/4}\;.
\end{align}
These observables correlate roughly linearly with corresponding moments of the initial eccentricity distributions: $\varepsilon_n\{2\}=\sqrt{\lr{\varepsilon_n^2}}$ and $\varepsilon_n\{4\} = (2\lr{\varepsilon_n^2}^2-\lr{\varepsilon_n^4})^{1/4}$. Crucially, while $v_n\{4\}$ captures primarily the average geometry component present in each event~\cite{ATLAS:2013xzf}, $v_n\{2\}$ is additionally sensitive to fluctuations of the geometry relative to the average, serving as complementary probes of the initial geometry.

{\bf Analysis.} 
The O+O and \dau data at $\snn=200$ GeV were collected in 2021 using minimum bias (MB) and high multiplicity (HM) triggers under low interaction rate conditions ($<30$ kHz) to reduce background. An upgraded time-projection chamber (TPC)~\cite{Anderson:2003ur,Wang:2017mbk,Chen:2024aom} extends the acceptance from $|\eta|<1$ in previous measurements~\cite{STAR:2022pfn,STAR:2023wmd} to $|\eta|<1.5$. Events were required to have a primary vertex within 30 cm of the TPC center along the beam axis and within 2 cm perpendicular to the beam axis. The final dataset includes approximately 500 million MB and 380 million HM O+O events, and 76 million MB and 70 million HM \dau events. 

Charged particles reconstructed in the TPC were selected within 0.2 $<\pT<$ 2.0 GeV/$c$ and $|\eta|<$ 1.5. The inverse of tracking efficiency $w(\pT,\eta,\phi)$ was obtained by embedding simulated charged particles into real data events~\cite{GEANT,FINE200176}. The number of charged particles $\nch$ was obtained by applying an efficiency correction for each track $i$ as $\nch = \sum_i w_i$.

The observables $v_n\{2\}$ and $v_n\{4\}$ require measuring moments of $p(v_n)$ (Eq.~\eqref{eq:1}): $\lr{v_n^2}$ and $\lr{v_n^4}$. They are obtained using two- and four-particle correlations for events in narrow $\nch$ ranges~\cite{ATLAS:2017rtr}:
\small{\begin{align}\nonumber
\lr{v_n^2}^{\mathrm{obs}} & =  \lr{ \frac{\sum_{i\neq j}w_iw_j \cos (n(\phi_i-\phi_j))}{\sum_{i\neq j}w_iw_j}}_{\mathrm{evt}}\;,\\\label{eq:2}
\lr{v_n^4}^{\mathrm{obs}} & =  \lr{ \frac{\sum_{i\neq j \neq k \neq l}w_iw_jw_kw_l \! \cos\!(n\!(\phi_i\!+\!\phi_j\!-\!\phi_k\!-\!\phi_l\!)\!)}{\sum_{i\neq j\neq k \neq l}w_iw_jw_kw_l}}_{\mathrm{evt}}\;.
\end{align}}\normalsize

These moments are contaminated by non-flow correlations from jet fragmentation and resonance decays~\cite{Jia:2017hbm}, with stronger impact for $v_n\{2\}$ than $v_n\{4\}$. To calculate $v_n\{2\}$, we require a large rapidity gap $|\Deta=\eta_i-\eta_j|>1.0$ on $\lr{v_n^2}^{\mathrm{obs}}$ to suppress non-flow correlations from near-side jets. Residual non-flow from away-side jets dominates at low charged particle multiplicity, scaling approximately as $1/\nch$. This away-side non-flow appears as a broad peak around $\Dphi=\phi_i-\phi_j\approx \pi$, contributing dominantly to $\lr{v_1^2}^{\mathrm{obs}}=\lr{\cos\Dphi}$ and modestly to $\lr{v_n^2}^{\mathrm{obs}}$~\cite{STAR:2023wmd}.

Following established procedures~\cite{STAR:2022pfn,STAR:2023wmd}, residual non-flow in $\lr{v_n^2}^{\mathrm{obs}}$ is assumed to scale as either $\lr{v_1^2}^{\mathrm{obs}}$ or $1/\nch$, and subtracted using low-multiplicity events (LM, $\nch < 20$):
\begin{equation}
\lr{v_n^2} =  \lr{v_n^2}^{\mathrm{obs}}- f\times \lr{v_n^2}^{\mathrm{obs,LM}} \;,
\label{eq:3}
\end{equation}
with $f = \lr{v_1^2}^{\mathrm{obs}}/\lr{v_1^2}^{\mathrm{obs,LM}}$ ($v_1^2$-scaling method) as default or $f = \nch^{\mathrm{LM}}/\nch$ ($\nch$-scaling) for estimating the non-flow systematics~\cite{STAR:2022pfn}. The flow coefficients at a single-particle level are then obtained as $v_n\{2\}= \sqrt{\lr{v_{n}^2}}$.
\begin{figure*}
\includegraphics[width=0.8\linewidth]{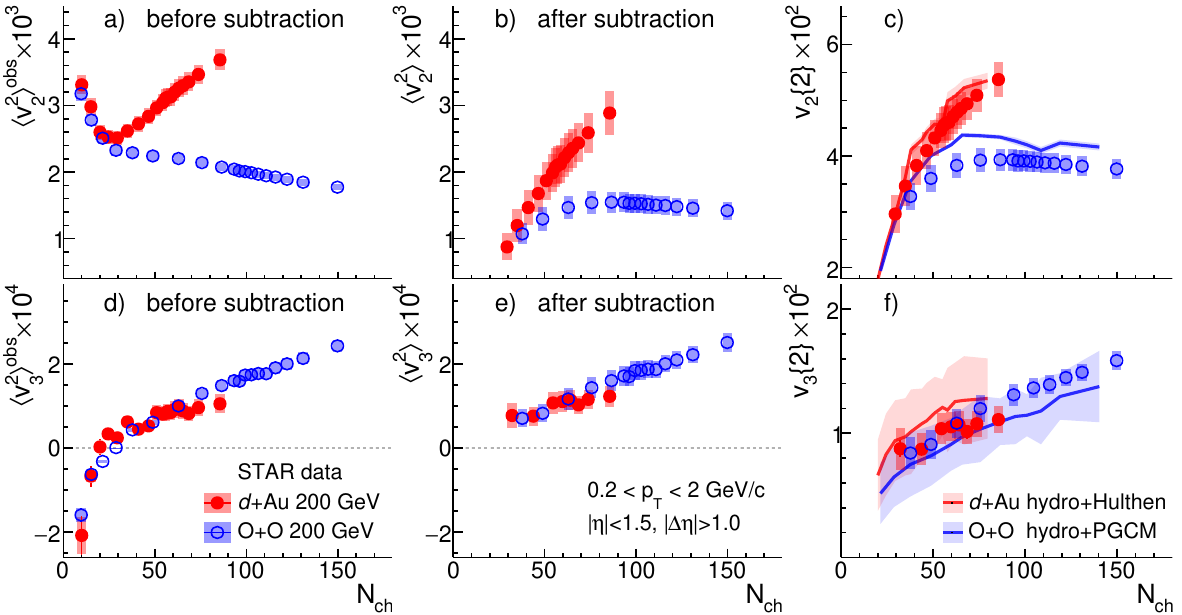}
\caption{\textbf{Flow coefficients from two-particle correlation method.} Raw correlators $\lr{v_n^2}^{\mathrm{obs}}$ (left), non-flow subtracted correlators $\lr{v_n^2}$ (middle), and flow coefficients $v_n\{2\}$ (right) versus $\nch$ for $n=2$ (top) and $n=3$ (bottom) in O+O and \dau collisions. Error bars denote statistical uncertainties, shaded bands denote systematic uncertainties. The $v_n\{2\}$ values are compared to hydrodynamic model predictions~\cite{Jahan:2025cbp}, whose error band represents model uncertainties.}
\label{fig:2}
\end{figure*}

The $v_2\{4\}$ measurements use the standard and subevent methods~\cite{Jia:2017hbm,ATLAS:2017rtr}. In the standard method, all multiplets in Eq.~\eqref{eq:2} are taken from $|\eta|<1.5$. In the two-subevent method, the multiplets are obtained by combining particles from two $\eta$ ranges: $-1.5<\eta_a<-0.1$ and $0.1<\eta_{b}<1.5$. In the three-subevent method, the multiplets are obtained by combining particles in three distinct $\eta$ ranges: $-1.5<\eta_a<-0.5$, $-0.5<\eta_b<0.5$, and $0.5<\eta_c<1.5$. The detailed implementation of these methods is described in Ref.~\cite{ATLAS:2017rtr,STAR:2025vbp}. In \oo collisions, two-subevent method is used as the default, with differences between two- and three-subevent methods assigned as systematic uncertainty. In \dau collisions, the much larger $v_2\{4\}$ signal compared to O+O allows consistent results between full-event and subevent methods. So the full-event method is used as default for superior statistical precision, and the differences between the full-event and subevent methods are assigned as non-flow uncertainty.

Systematic uncertainties for $v_{2}\{2\}$, $v_{3}\{2\}$, and $v_2\{4\}$ arise from three sources. 1) Event and track selection: varying the range of primary vertex positions and requirements on the distance of closest approach and the number of TPC space points associated with the tracks. The changes are less than 2\% for $v_2\{2\}$, 5\% for $v_3\{2\}$, and 10\% for $v_2\{4\}$. 2) Trigger bias: evaluated by comparing results from MB and HM triggers. The differences are less than 2\% for $v_2\{2\}$, 5\% for $v_3\{2\}$, and 5\% for $v_2\{4\}$. 3) Residual non-flow: non-flow for $v_2\{2\}$ and $v_3\{2\}$ is estimated by comparing results from the two subtraction methods and varying $\Delta\eta$ gap from $|\Delta\eta|>1.0$ to $|\Delta\eta|>1.2$, as well as same-charge versus opposite-charge particle pairs. The choice of the LM events is also varied from the default range of $13<\nch<18$ to $7<\nch<13$. The change in results is less than 10\% in central collisions and up to 20\% in more peripheral collisions. The non-flow uncertainties for $v_2\{4\}$ are found to be less than 10\%. Most uncertainties are strongly correlated with $\nch$, and hence won't affect the trends of flow observables as a function of $\nch$. These sources are combined in quadrature to obtain the total systematic uncertainties.

The $\varepsilon_n$ in \dau and O+O collisions are calculated using the standard PHOBOS Glauber model~\cite{Alver:2008aq} as described in Ref.~\cite{Huang:2025cjm}. Nucleon configurations of deuteron are sampled from the Hulthen function~\cite{Hulthen:1957}, while those for oxygen are generated from three {\it ab initio} nucleon configuration models: nuclear lattice effective theory (NLEFT)~\cite{Lu:2018bat}, projected generator coordinate method (PGCM)~\cite{Frosini:2021ddm}, and auxiliary field diffusion Monte Carlo (AFDMC)~\cite{Lonardoni:2017hgs}. These models differ in the nuclear forces and methods used for generating nucleon configurations. Subnucleonic fluctuations are implemented by modeling each nucleon as three constituent quarks~\cite{Loizides:2016djv}. The $\varepsilon_n$ are then calculated from positions of participating nucleons (nucleon-Glauber model) or participating quarks (quark-Glauber model). 

To match the measured multiplicity distribution $p(\nch)$, we assume $\nch$ in each event is proportional to the number of sources -- either participating nucleons or participating quarks~\cite{STAR:2023wmd}. The multiplicity for each source is sampled from a negative binomial distribution whose parameters are tuned to match experimental data. The corresponding eccentricities are calculated as a function of $\nch$. Following procedures outlined in Ref.~\cite{STAR:2023wmd}, the fitted $p(\nch)$ from the nucleon-Glauber model is used to define event centrality (see appendix). 

{\bf Results.} 
Figure~\ref{fig:2} shows results from the two-particle correlation method. The raw correlators $\lr{v_n^2}^{\mathrm{obs}}$ in the left panels reveal intricate interplay between genuine collective flow and non-flow correlations. At low $\nch$, non-flow from away-side jet fragmentation dominates (positive for $n=2$, negative for $n=3$). At higher $\nch$, genuine flow dominates the $\nch$-dependent behavior. This remarkable interplay produces the non-monotonic $\nch$ dependence of $\lr{v_2^2}^{\mathrm{obs}}$ in $\dau$ collisions and the sign change of $\lr{v_3^2}^{\mathrm{obs}}$ versus $\nch$ in both systems.

After non-flow subtraction, the resulting $\lr{v_n^2}$ in the middle panels reveals distinct ordering between systems. Namely, $\lr{v_{2,d\mathrm{Au}}^2}$ increases rapidly with $\nch$, while $\lr{v_{2,\mathrm{OO}}^2}$ rises to a maximum around $\nch\sim 75$ then decreases slightly. This rise-and-fall behavior reflects the change in average elliptic geometry from mid-central to central O+O collisions. In contrast, $\lr{v_3^2}$ shows a more gradual increase with $\nch$ in both systems, characteristic of a fluctuation-driven origin.

The corresponding single-particle flow coefficients $v_n\{2\}=\sqrt{\lr{v_n^2}}$ are shown in the right panels. The $v_2\{2\}$ values reach nearly 0.06 in \dau collisions, about 40\% larger than in \oo collisions, whereas $v_3\{2\}$ values agree between systems and reach up to 0.015 in central O+O collisions. These distinct trends strongly suggest that the observed flow reflects collective response to the initial geometry.

\begin{figure}
\includegraphics[width=1.0\linewidth]{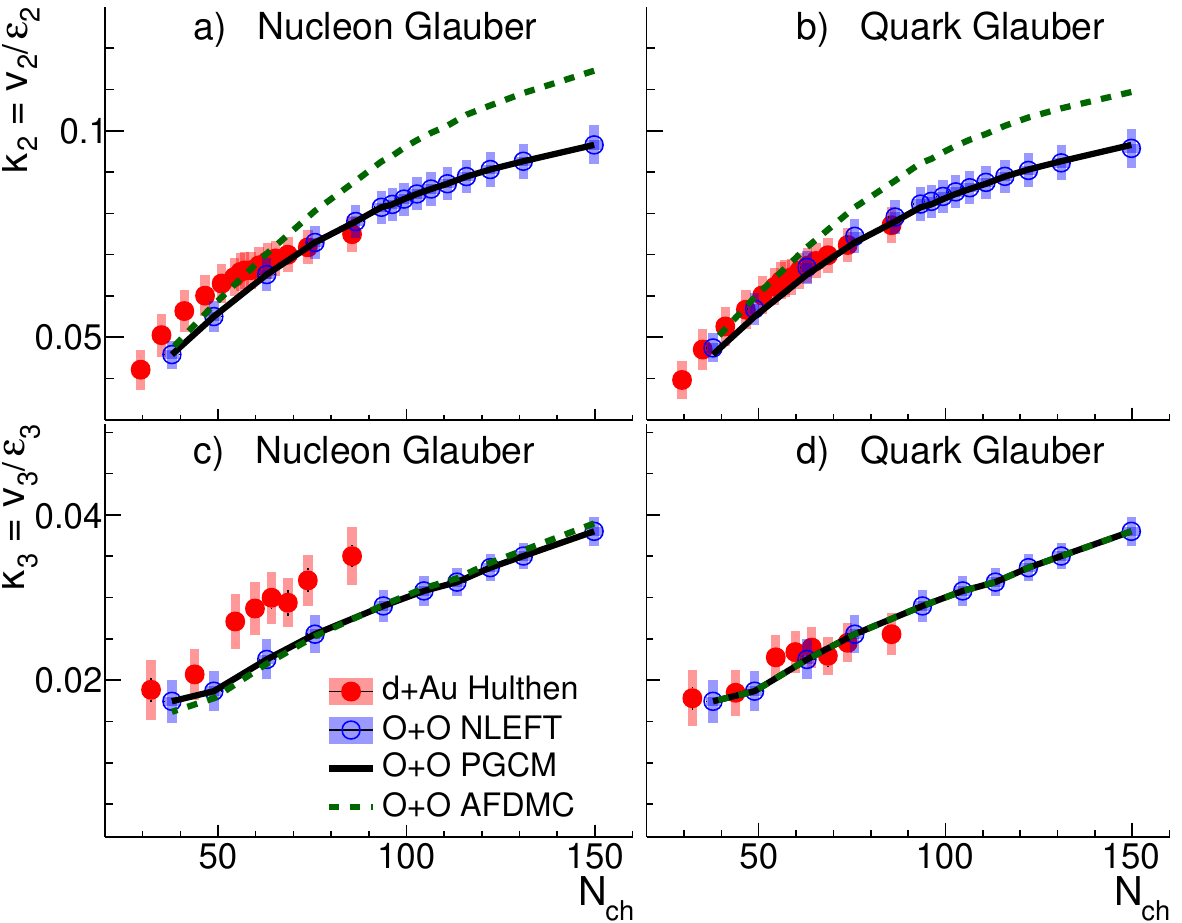}
\caption{\textbf{Impact of nuclear and subnucleonic structures.} The final-state response coefficients $k_n = v_n\{2\}/\varepsilon_n\{2\}$ from nucleon-Glauber (left) and quark-Glauber (right) models versus $\nch$ for $n=2$ (top) and $n=3$ (bottom) in \dau and O+O collisions. The $\varepsilon_n^{\mathrm{OO}}$ are calculated using nucleon configurations of $^{16}$O obtained from three models: NLEFT, PGCM, and AFDMC (see text). Error bars and shaded bands represent statistical and systematic uncertainties, respectively. The $v_n\{2\}$ data uncertainties, common to all models, are shown only for NLEFT for clarity.}
\label{fig:3}
\end{figure}

Figure~\eqref{fig:3} shows the response coefficients $k_n = v_n\{2\}/\varepsilon_n\{2\}$ with $\varepsilon_n\{2\}$ obtained from Glauber models, which quantify the medium's efficiency in converting initial shape anisotropies to final momentum anisotropies. We observe approximate scaling of $k_n$ between the two systems. The agreement of $k_n$ between systems is imperfect in the nucleon-Glauber model (panels a and c), but improves when including subnucleonic fluctuations (panels b and d). This demonstrates that symmetric-asymmetric system comparison is sensitive to both nuclear and subnucleonic structure: $v_2$ reflects mainly nucleon configurations, while $v_3$ is more sensitive to subnucleonic fluctuations.

At large $\nch$, $k_2^{\mathrm{OO}}$ from three {\it ab initio} structure models shows divergence, caused by the ordering of eccentricity: $\varepsilon_2^{\mathrm{NLEFT}}\approx\varepsilon_2^{\mathrm{PGCM}}>\varepsilon_2^{\mathrm{AFDMC}}$~\cite{Zhang:2024vkh}. Clearly, $k_2^{\mathrm{OO}}$ based on AFDMC exhibits significant deviation from $k_2^{d\mathrm{Au}}$, whereas those from the other two models agree with $k_2^{d\mathrm{Au}}$. In contrast, $k_3^{\mathrm{OO}}$ values agree among the three model inputs, consistent with random fluctuation dominance for triangularity.

The final-state responses are investigated using a state-of-the-art hydrodynamic model~\cite{Jahan:2025cbp} that includes appropriate initial geometries and medium properties. The initial condition is generated using nucleon configuration for $^{16}$O from PGCM~\cite{Frosini:2021ddm} with four hotspots per nucleon and specific longitudinal structure. This initial condition is then interfaced to MUSIC hydrodynamic model for QGP evolution, followed by UrQMD for hadronic evolution. The QGP parameters are provided by Bayesian inference of Au+Au collision data. However, this procedure significantly underestimates $\varepsilon_3^{d\mathrm{Au}}$, hence initial state parameters were revised slightly to calculate $\varepsilon_3$ and $v_3$. As shown in the right panels of Fig.~\ref{fig:2}, this model quantitatively reproduces the $\nch$ dependence of $v_2$ and $v_3$ in both systems, implying that the complex flow patterns can be successfully described assuming the creation of a QGP with properties similar to those in large systems.

\begin{figure}
\includegraphics[width=1.0\linewidth]{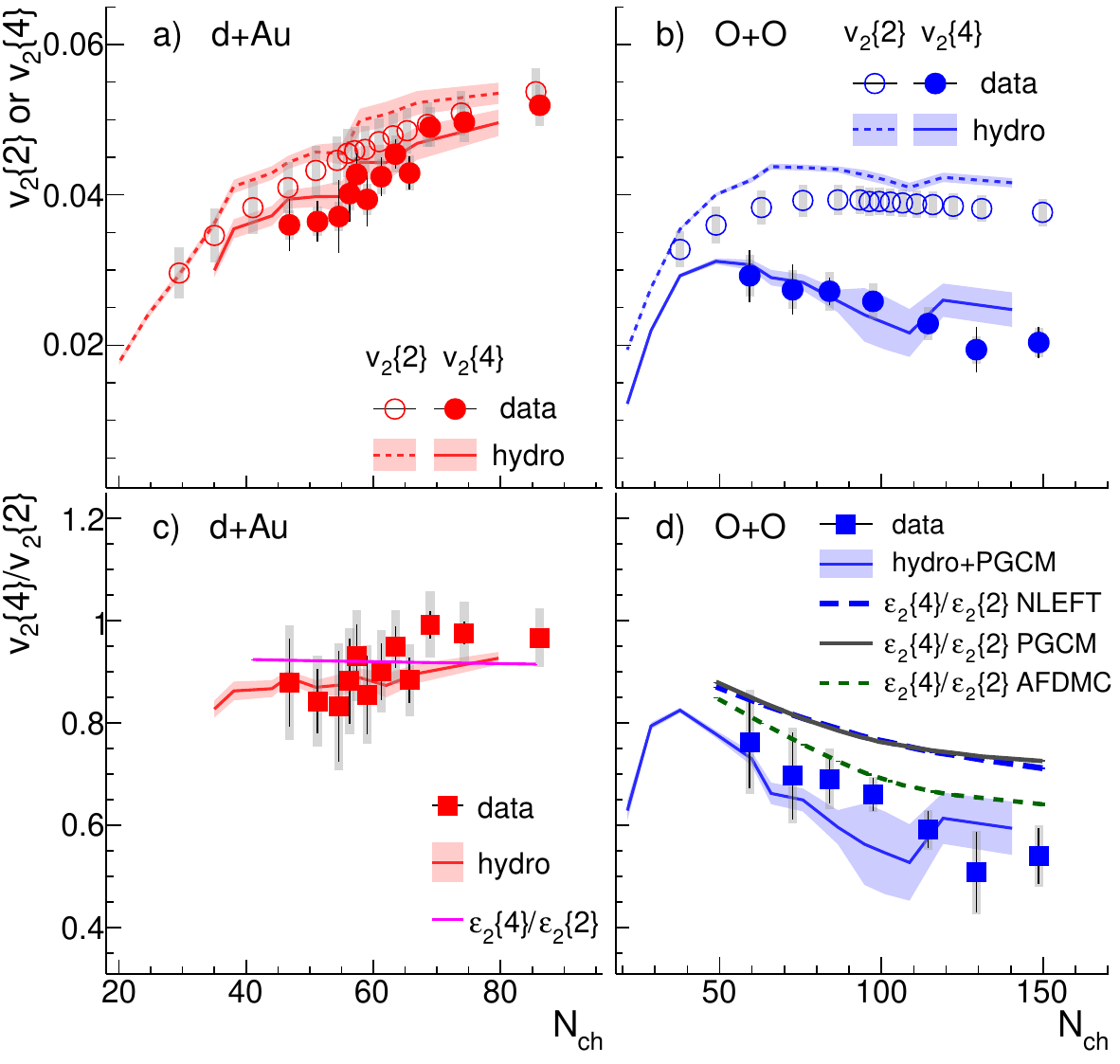}
\caption{\textbf{Contrasting elliptic flow fluctuations.} Flow coefficients $v_2\{2\}$ and $v_2\{4\}$ (top) and their ratio $v_2\{4\}/v_2\{2\}$ (bottom) versus $\nch$ in \dau (left) and O+O (right) collisions. Error bars and shaded bars denote statistical and systematic uncertainties, respectively. Results are compared with hydrodynamic model calculations~\cite{Jahan:2025cbp} and, for O+O, with quark-Glauber calculations using three different {\it ab initio} nucleon configuration theories.}
\label{fig:4}       
\end{figure}

Figure~\ref{fig:4} presents measurements of elliptic flow fluctuations using $v_2\{4\}$ and $v_2\{2\}$. The pronounced elliptic shape in $d$+Au collisions enhances the average geometry component, leading to $\varepsilon_2\{4\}\approx \varepsilon_2\{2\}$. In contrast, O+O collisions have a much smaller average geometry component, resulting in $\varepsilon_2\{4\} < \varepsilon_2\{2\}$. In a hydrodynamic scenario where $v_2\propto \varepsilon_2$, these relations are expected to survive to the final state. Indeed, we find $v_2\{4\}/v_2\{2\} \approx 0.9$ in $d$+Au collisions, while in O+O collisions this ratio decreases with $\nch$ from 0.8 to 0.5. We observe that $v_2\{4\}/v_2\{2\}$ from the hydrodynamic model is systematically smaller than $\varepsilon_2\{4\}/\varepsilon_2\{2\}$, due to residual non-linear effects in $k_2$ that affects $v_2\{4\}$ more strongly~\cite{Noronha-Hostler:2015dbi}. However, these non-linear effects should be independent of the initial geometry models, hence the modest sensitivity of $\varepsilon_2\{4\}/\varepsilon_2\{2\}$ to {\it ab initio} nucleon configuration models observed in Fig.~\ref{fig:4}d are expected to survive to $v_2\{4\}/v_2\{2\}$.

The hydrodynamic model reproduces remarkably well the substantial ordering of $v_2\{2\}$ and $v_2\{4\}$, as well as the ratio $v_2\{4\}/v_2\{2\}$, over the full $\nch$ range in both systems. This agreement establishes unambiguously the geometrical origin of $v_2$, which differs markedly between symmetric and asymmetric collisions. These results also rule out scenarios in which initial-state momentum correlations play a significant role in the measured flow for these small systems.

{\bf Summary.}
We have demonstrated the ability to engineer the shape of the medium by comparing $v_2$ and $v_3$ in asymmetric $d$+Au and symmetric O+O collisions at $\snn =200$ GeV. These systems produce mediums of comparable sizes and $\varepsilon_3$, but with very different $\varepsilon_2$ and fluctuation patterns. These differences serve as a powerful lever arm to test the impacts of nucleon configuration and subnucleonic fluctuations on initial geometries and final-state hydrodynamic responses.

Flow coefficients obtained using two-particle correlation method $v_2\{2\}$ and $v_3\{2\}$, and four-particle correlation method $v_2\{4\}$ reveal several key insights. We find $v_2^{d\mathrm{Au}}\{2\}> v_2^{\mathrm{OO}}\{2\}$ at large $\nch$, reflecting the large intrinsic elliptic shape of the deuteron. This intrinsic shape suppresses event-by-event fluctuations of $v_2$, leading to $v_2^{d\mathrm{Au}}\{4\}\approx 0.95 v_2^{d\mathrm{Au}}\{2\}$, in contrast to $v_2^{\mathrm{OO}}\{4\}\approx 0.5 v_2^{\mathrm{OO}}\{2\}$ in central O+O collisions. In contrast, we observe $v_3^{d\mathrm{Au}}\{2\}\approx v_3^{\mathrm{OO}}\{2\}$ across the full $\nch$ range, consistent with the fluctuation-driven scenario in the quark-Glauber model.

Flow coefficients correlate strongly with their corresponding initial state eccentricities in the two systems. Nucleon configurations of the deuteron and $^{16}$O play a dominant role in determining $\varepsilon_2$ and $\varepsilon_3$, and consequently $v_2$ and $v_3$. Furthermore, including subnucleonic fluctuations in the deuteron improves the scaling between $v_2$ and $\varepsilon_2$, and more strikingly between $v_3$ and $\varepsilon_3$, highlighting the necessity for considering such effects. Finally, $\varepsilon_2$ and its event-by-event fluctuations are sensitive to different models of the nucleon configuration of $^{16}$O, which may impact the measured $v_2$ and its fluctuations.

The measured flow coefficients are quantitatively reproduced by a state-of-the-art hydrodynamic model tuned to Au+Au data. These results suggest that collective flow in small systems originates from hydrodynamic response to initial geometries, consistent with the formation of QGP droplets whose transport properties are similar to those observed in large collision systems.

During the final preparation of this Letter, ATLAS, ALICE and CMS Collaborations submitted $v_2$ and $v_3$ results in O+O and Ne+Ne collisions at $\snn=5.36$ TeV~\cite{ATLAS:2025nnt,ALICE:2025luc,CMS:2025tga}. The availability of O+O data at both RHIC and LHC opens up an exciting opportunity to study origin of flow across a wide energy range. 

{\it Acknowledgments} 
We thank the RHIC Operations Group and SDCC at BNL, the NERSC Center at LBNL, and the Open Science Grid consortium for providing resources and support. This work was supported in part by the Office of Nuclear Physics within the U.S. DOE Office of Science, the U.S. National Science Foundation, National Natural Science Foundation of China, Chinese Academy of Science, the Ministry of Science and Technology of China and the Chinese Ministry of Education, NSTC Taipei, the National Research Foundation of Korea, Czech Science Foundation and Ministry of Education, Youth and Sports of the Czech Republic, Hungarian National Research, Development and Innovation Office, New National Excellency Programme of the Hungarian Ministry of Human Capacities, Department of Atomic Energy and Department of Science and Technology of the Government of India, the National Science Centre and WUT ID-UB of Poland, the Ministry of Science, Education and Sports of the Republic of Croatia, German Bundesministerium f\"ur Bildung, Wissenschaft, Forschung and Technologie (BMBF), Helmholtz Association, Ministry of Education, Culture, Sports, Science, and Technology (MEXT), and Japan Society for the Promotion of Science (JSPS). We thank Chun Shen for providing hydrodynamic model calculations.

\section*{Appendix}
\begin{figure}
\includegraphics[width=1.0\linewidth]{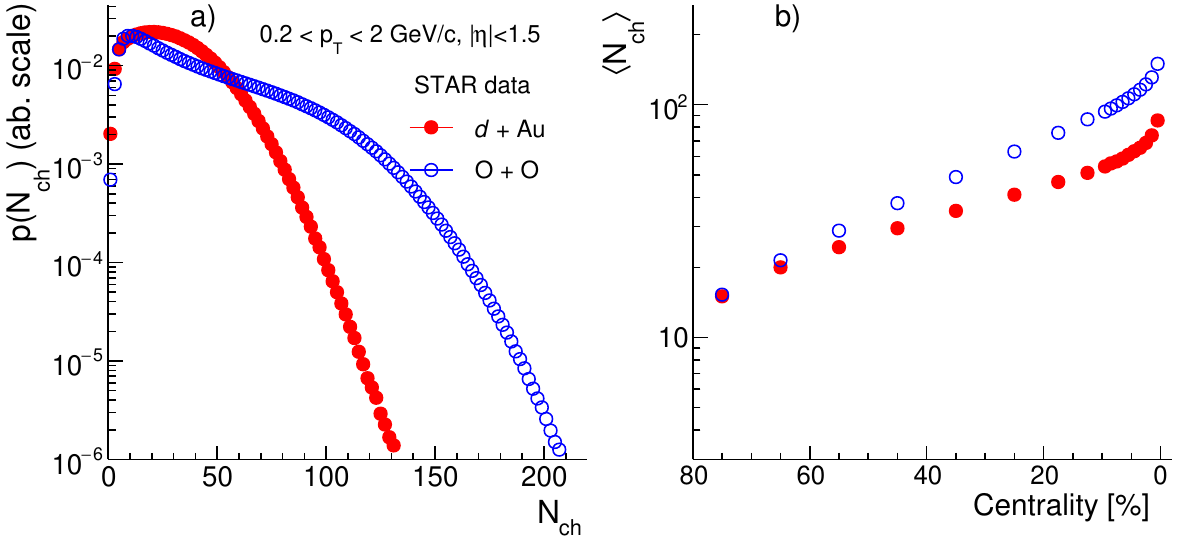}
\caption{ The distribution of $\nch$ (left) and $\lr{\nch}$ vs centrality (right) in \dau and O+O collisions.}
\label{fig:5}       
\end{figure}

This analysis is performed as a function of efficiency corrected charged particle multiplicity in 0.2 $<\pT<$ 2.0 GeV/$c$ and $|\eta|<$ 1.5, $\nch$. This distribution $p(\nch)$ is broader in O+O collisions than \dau collisions, as shown in Fig.~\ref{fig:5}a. A nucleon Glauber model fit to this distribution~\cite{STAR:2023wmd} is then used to determine the relation between $\nch$ and event centrality as shown in Fig.~\ref{fig:5}b.

\bibliography{ref}
\end{document}